\documentclass{cjaa}

\usepackage{graphicx,times,color}
\input{epsf.sty}
\input{psfig.sty}
\begin{document}

   \title{Determination of Topology Skeleton of Magnetic Fields in a Solar Active
   Region$^*$
\footnotetext{$*$ Supported by the National Natural Science
Foundation of China.} }

\volnopage{Vol.\ 8 (2008), No.,~ 000--000}
   \setcounter{page}{1}

   \author{Hui Zhao
        \inst{1}
   \and Jing-Xiu Wang
        \inst{1}
   \and Jun Zhang
        \inst{1}
   \and Chi-Jie Xiao
        \inst{1}
   \and  Hai-Min Wang
      \inst{2}
          }

   \institute{National Astronomical Observatories, Chinese Academy of Sciences,
             Beijing 100012; {\it wangjx@bao.ac.cn}\\
     \and
     Big Bear Solar Observatory, 40386 North Shore Lane, Big Bear City,
        CA 92314, USA \\
   \vs\no
   {\small Received 2007 ; accepted 2007 }
}

\abstract{ The knowledge of magnetic topology is the key to
understand magnetic energy release in astrophysics. Based on
observed vector magnetograms, we have determined three-dimensional
(3D) topology skeleton of the magnetic fields in active region NOAA
10720. The skeleton consists of six 3D magnetic nulls and a network
of corresponding spines, fans, and null-null lines. For the first
time, we have identified a spiral magnetic null in Sun's corona. The
magnetic lines of force twisted around the spine of the null,
forming a `magnetic wreath' with excess of free magnetic energy and
resembling observed brightening structures at extra-ultraviolet
(EUV) wavebands. We found clear evidence of topology eruptions which
are referred to as the catastrophic changes of topology skeleton
associated with a coronal mass ejection (CME) and an explosive X-ray
flare. These results shed new lights in exploring the structural
complexity and its role in explosive magnetic activity. In solar
astrophysics and space science, the concept of flux rope has been
widely used in modelling explosive magnetic activity, although their
observational identity is obscure or, at least, lacking of necessary
details. The current work suggests that the magnetic wreath
associated with the 3D spiral null is likely an important class of
the physical entity of flux ropes.
   }

   \authorrunning{H. Zhao, J.-X. Wang, J. Zhang et al.}
   \titlerunning{Determination of Topology Skeleton}

   \maketitle

\section{Introduction}
\label{sect:intro}

The self-closure and frozen-in condition of magnetic fields in
astrophysical and space plasmas make the electromagnetic
interaction extremely complicated in the sense that the magnetized
plasma is divided into distinct topologies. The explosive release
of stored magnetic energy, which appears presumably in the
topology interface, can not be realized without topology collapse.
Solar active regions (ARs) represent a typical example of
electromagnetic interaction in astrophysics. They are
characterized by strong magnetic fields with complex topology,
serving as a paradigm in astrophysics and space science. Since
late 1940s it has been recognized that under certain conditions
the breakdown of magnetic topology could produce solar flares and
cosmic ray particles (Giowanelli 1946; Sweet 1958). To explore the
magnetic skeleton in an AR, which comprises magnetic nulls
(neutral points) and a network of spines ($\gamma$ lines) and fans
($\Sigma$ surfaces) (Lau \& Finn 1990; Parnell et al. 1996), is
fundamental in understanding explosive solar activity (Priest et
al. 1997). This, in turn, will guide to a thorough understanding
of ubiquitous magnetic activity in the universe.

The 3D topology skeleton of Sun's magnetic fields, which is
composed of nulls, spines, fans and separators (Priest et al.
1997), has only been described either by simple analytical models
(Lau 1993; Brown \& Priest 2001; Parnell 2007), or by theoretical
calculations with `magnetic charges' or `dipoles' to approximate
the observed magnetic fields on solar surface (Seehafer 1986;
Gorbachev \& Somov 1988; D\'{e}moulin et al. 1992; Longcope \&
Klapper 2002; Longcope 2005). Comparing brightness observations
with the above conceptual models, Filippov (1999), Aulanier et al.
(2000), Fletcher et al. (2001), Maia et al. (2003) and Li et al.
(2006) identified 3D nulls in the corona. These approaches have
demonstrated that there, indeed, exist 3D magnetic nulls with
spines and fans in the corona, forming a skeleton of magnetic
topology. In addition, the magnetic energy release in solar flares
is closely associated with these topology structures. However, as
too many simplifications are adopted, e.g., the representation of
the observed fields by discrete `magnetic charges' or `dipoles'
and the assumption of the current-free (potential) fields, we are
far from being able to evaluate current status of our knowledge
about the real magnetic skeleton, let alone the nature of 3D
magnetic reconnection in the solar atmosphere (Parnell 2007). Only
recently, 3D magnetic nulls are identified in geomagnetotail based
on the \emph{in situ} measurements with Cluster spacecraft (Xiao
et al. 2006). On the other hand, separatrices can also be
approached by discontinuities in the footpoint mapping which is
defined in a continuous magnetic boundary (Low 1987; Low \&
Wolfson 1988). This discontinuities ascribe coronal nulls or bald
patches (Seehafer 1986; Titov et al. 1993; Bungey et al. 1996).
Some authors (Priest \& D\'{e}moulin 1995; D\'{e}moulin et al.
1996; Titov et al. 2002) give out measurement of mapping
distortion or squashing to accurately locate the so-called
quasi-separatrix layer (QSL). The QSL method is not a direct way
to determine the separatrices. So it would take more time even
than magnetic charge topology (MCT) method (Longcope 2005) to
approach the same singular feature. From the footpoint mapping of
magnetic boundary we also can not get clear skeleton as the
methods depart from the nulls. But to study the features in
photosphere or lower coronal, it is still a useful tool for its
precise measurement of the discontinuities of the field lines.

The solid advances in the measurements of solar vector magnetic
fields in recent years have open a possibility to observationally
determine the topology skeleton in the Sun's active corona. In
this paper, based on high quality and high cadence vector
magnetograms and theoretical calculations, we have completely
determined the topology skeleton of magnetic fields for a solar
AR, NOAA 10720. The determined topology skeleton and its temporal
evolution are approached, then, the explosive magnetic energy
release in a major solar flare/CME event is understood with the
knowledge of topology skeleton.

This paper is organized as followings: Section~2 is devoted to a
description of observations of vector magnetic fields in AR 10720;
Section~3 is a brief introduction of the methods used in this work
to determine the topology skeleton; Section~4 is main results of the
study on the determined topology skeleton and its temporal evolution
which led to the major flare and CME activity; In Section~5, we
present a particular discussion on the possible uncertainty of the
results; Conclusions are finally summarized in the last section.

\section{Observations of vector magnetic fields in AR 10720}
\label{sect:Obs}

NOAA AR 10720 is a super AR which has created the largest and
hardest proton flux ($>$100\,MeV) since 1989 and disastrous
condition in space weather. It rotated onto the solar disk as a
simple beta magnetic sunspot on January 10 and ended as a large,
magnetic complex sunspot region on January 22, 2005. It grew rapidly
and showed impressive activity while it transited the solar disk.
During January 14--21 it produced 5 X-class flares and 18 M-class
flares. Vector magnetic fields were mapped using Huairou Solar
Observing Station (HSOS) vector-magnetograph (Ai and Hu 1986; Wang
et al. 1996) from January 12 to 20, which provide clues in
understanding the detailed magnetic evolution associated with major
flares. Big Bear Solar Observatory (BBSO) obtained vector
magnetograms in January 14--15 with high cadence of 1--2 minutes and
high resolution around 1--2$''$. The vector field evolution, thusly,
could be traced in a round-the-clock way from January 14--16.

The BBSO vector magnetograms were obtained by Digital Vector
Magnetograph (DVMG) system, covering an area of about 300$''$
$\times$ 300$''$. It consists of a  1/4 \AA~ band pass
birefringent filter, an SMD 1024 $\times$ 1024 12-bit CCD camera
and polarization analyzer. Each data set consists of four images:
a 6103 \AA~filtergram (Stokes-I), a line-of-sight magnetogram
(Stokes-V) and the transverse magnetogram (Stokes-U and -Q). We
rebin the camera to the 512 $\times$ 512 mode to increase the
sensitivity of the magnetograms.

The polarization analyzer includes two nematic liquid crystal
variable retarders, whose retardance can be controlled with an
applied voltage, to select a particular polarization state (6103
\AA~ filtergram, Stokes V, Q, or U) by converting the desired input
polarization set into an orthogonal set of linear polarizations. A
single ferroelectric liquid crystal, which is a fixed retarder whose
rotation angle can be selected to be either 0 or 45 degree, acts as
the system's fast modulator. This ferroelectric crystal, working
with a fixed linear polarizer, is used to select one of the
orthogonal linear polarization components. Light is then fed through
the birefringent filter, and finally imaged onto the CCD camera. The
exposure is typically 30 ms, and images are taken at a rate of
approximately 12 frames s$^{-1}$.

A distinct characteristic of the AR evolution is the appearance of
sheared emerging flux regions (EFRs) which were current-carrying.
They emerged successively and grew along the main magnetic neutral
line of the AR. In Figure~1, a time sequence of vector magnetograms,
obtained from Huairou Solar Observing Station (HSOS) of National
Astronomical Observatories of Chinese Academy of Sciences, is shown.
These EFRs make the AR grown rapidly since January 13. On the figure
four EFRs are marked by brackets. A very key characteristic of these
EFRs is the appearance of a bundle of enhanced transverse fields
which are connecting the growing magnetic flux of opposite polarity.
The emergence of the four EFRs can be traced since January 13 until
January 18. It is interesting that all these four EFRs were growing
along the main magnetic neutral line, and separating in roughly the
opposite direction perpendicular to the the main magnetic axis of
the AR. It is noticed that the EFR marked close to the center of the
magnetogram at 01:20 UT of January 15 showed opposite sense of
twisting, i.e., opposite sign of magnetic shear or current helicity
in comparing with other EFRs. The co-existence of opposite helicity
in new EFRs made the 3D magnetic configuration more complicated.
Later from January 16, another new EFR appeared in the area of
negative flux to the north of the magnetic neutral line (see the
growing positive magnetic flux).
\begin{figure}
   \centering
   \includegraphics[scale=0.6]{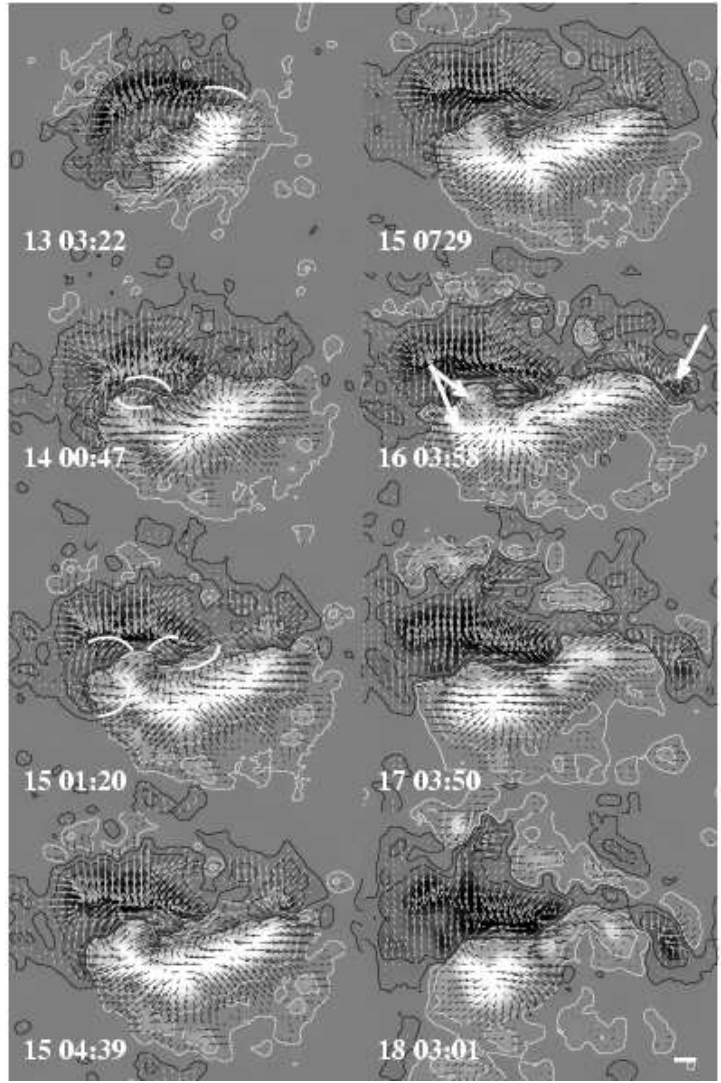}
   \caption{ Time sequence of Huairou vector magnetograms. The line of sight
   component is shown by images scaled from --1500 to 1500\,G and contours with
   levels of $\pm 50, 250, 500, 1000, 1500$\,G. The transverse components are
   shown by short arrows with length proportional to the field strength and
   direction indicated by the arrows. Four key EFRs are marked in the
   magnetograms by brackets. Each EFR is characterized by a bundle of enhanced
   transverse fields and two growing footpoints of opposite polarity. At
   03:08 UT of January 16, three thick arrows indicate the concentrated
   magnetic flux of opposite polarity which come from the successive EFRs
   from the interval of January 13 to 16. The scale bar in the lower-bottom
   corner denotes 20.0 arcsec.}
   \label{Fig:magnetogram}
   \end{figure}

The effects of these EFRs are presented in several perspectives: (1)
The magnetic neutral line was largely elongated. From 00:32 UT of
January 13 to 16:51 UT of January 15, the neutral line was elongated
from 120 Mm to 260 Mm; (2) Interleaving opposite polarities and
multiple neutral lines appeared in the center of the AR; (3) Strong
magnetic shear developed along this elongated neutral line in
roughly the East-West direction. The net results of these EFRs
appear to be equivalent to a widely separated `EFR' as indicated by
thick arrows in the magnetogram at 03:58 UT of January 16. Overall
speaking, the opposite polarities of newly emerged flux separated at
a velocities of 0.5--0.7\,km~s$^{-1}$ in the observed interval.

The central meridian passage of AR 10720 took place On January 15.
An X-ray flare with significance of X2.6 appeared at 22:25 UT in the
AR. This flare is one of a significant flares associated with a
full-halo CME and serves as a typical active event for careful
exemplification. From 16:51 to 23:47 UT BBSO obtained 262 sets of
high quality vector magnetograms with 20 Gauss sensitivity in
line-of-sight magnetograms, and 150 Gauss in transverse magnetograms
(Spirocks 2005). These observations fully confirmed the evolutionary
characteristics of the vector magnetic fields in the AR which were
revealed by HSOS. In Figure~2, vector field evolution in 7-hour
interval is shown in details. The consistence in flux distribution
and field azimuth is impressive between HSOS and BBSO magnetograms.
The well-known 180 degree ambiguity in the observed field azimuth is
resolved by potential and constant-$\alpha$ force-free field
assumptions. A few empirical corrections, based on the history of
flux emergence and the continuity in the observed azimuth, are made
after removing the ambiguity in the above objective ways.
\begin{figure}
   \centering
  \includegraphics[scale=0.6]{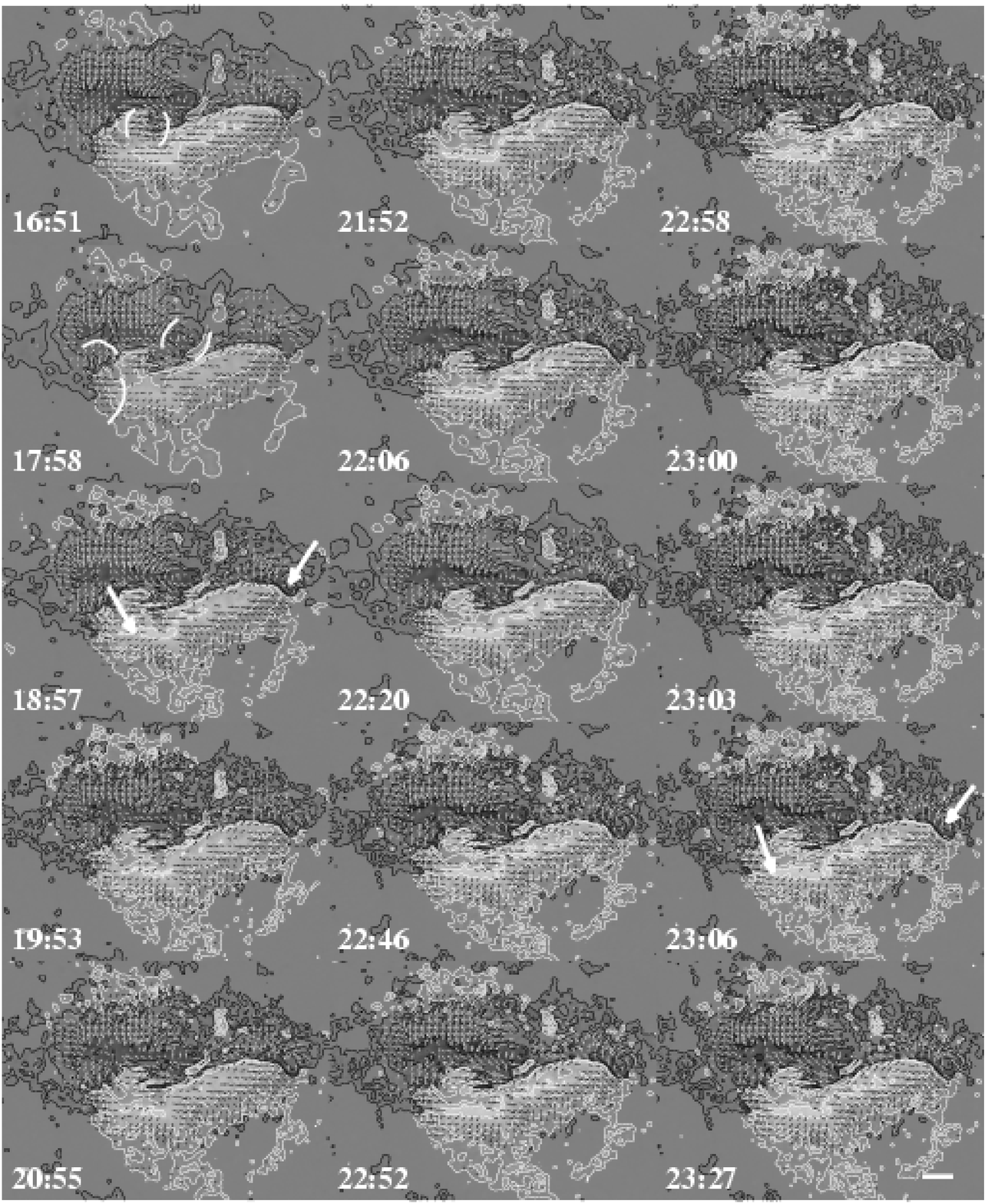}
   \caption{ Time sequence of BBSO magnetograms which are shown by the same
   style as in Figure 1. Three EFRs in continuous growing are marked in the
   magnetograms by brackets. Two sets of thick arrows indicate the concentrated
   magnetic flux of opposite polarity which come from the successive EFRs
   since January 13. Note, the opposite polarities of the equivalent EFR are
   in continuous growing and separating. The scale bar denotes 25.0 arcsec.}
   \label{Fig:magnetogram evolution}
   \end{figure}

There is no doubt that the topology skeleton in the magnetized
plasma of the AR is determined by the magnetic field distribution
and evolution in the photosphere. BBSO vector magnetograms were
obtained at much better seeing and with a cadence of 1--2 minutes.
In addition, the BBSO vector magnetograms covered the whole interval
of the major flare/CME on January 15 from pre-status to
full-recovery phases. Therefore the BBSO magnetograms provide a rare
opportunity to observationally determine the magnetic skeleton of a
solar AR, and to identify the characteristics of the skeleton
evolution in the course of explosive energy release in solar
activity.

\section{Methods}
\label{sect:method}

We use the observed vector magnetograms of BBSO as boundary
conditions, employing a quasi-linear force-free code (Wang et al.
2001), we reconstruct the 3D magnetic fields of AR 10720 first.
For the magnetograms which we selected are very near to the center
of the solar disk, we can use them directly without solar rotation
correction. Due to the influence of the stray light, there exist
areas of polarization saturation, which appeared as `big holes' in
the middle of the strong sunspots in longitudinal magnetograms. We
use the MDI longitudinal magnetogram at the nearest time to
fulfill these `holes'.

The size of extrapolation box is 250 $\times$ 250 $\times$ 150 in
bins. Each bin corresponds to 2.1$''$. The observed magnetogram
embed in the middle of 150 $\times$ 150 as shown in Figure~4 of this
paper.

We search for nulls in every cell of the box with Poincar\'{e} index
(Greene 1992; Zhao et al. 2005) after we obtain the 3D vector fields
at each node in the AR. This method in 2D conditions have been used
on vector magnetograms by Wang \& Wang (1996). In 3D condition, we
get at first the magnetic vectors in eight vertex of each cell, then
we normalize these vectors and translate these images of vectors to
a unit sphere. The cubic surface of each cell can naturally be
separated into 12 triangles (see fig.~2 in Zhao et al. 2005), every
triangle map to a spherical triangle in the unit sphere. By
integrating these twelve spherical triangles and dividing the
integral by $4\pi$, we obtain the Poincar\'{e} index in practice. A
little different from 2D conditions (Wang \& Wang 1996), if a cell
include a null point, the index should be equal to 1 or --1, if not,
0. The nulls whose initial positions are outside the middle 150
$\times$ 150 $\times$ 50 bins or lower than 1 grid are ignored, for
extrapolation in those regions is not very credible, or foot points
of the field lines through this region overstep the observed view.
We extend the extrapolation scale for the sake of pursuing some
nulls' evolution.

\begin{figure}
   \centering
   \includegraphics[scale=0.32]{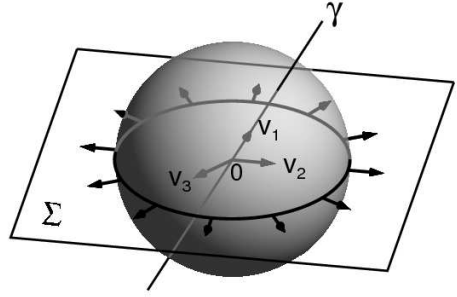}
   \caption{Illustrate the structures near to a (B type) null. A unit sphere will
   wrap a cell which include a null. The $\gamma$ line and the $\Sigma$ surface of
   the null intersect at it. They cross the sphere at two points and a circle, in which
   the vectors parallel to the directions of the $\gamma$ line and the $\Sigma$ surface,
    respectively. These directions are determined by the three eigenvectors, $\textbf{v}_{1}$,
   $\textbf{v}_{2}$, $\textbf{v}_{3}$, of the matrix $\delta\textbf{B}$.}
   \label{Fig:sphere}
   \end{figure}

In order to draw out the magnetic skeleton above the active region,
we calculate the field's Jacobian matrix $\delta \textbf{B}$ in each
null-point which has been identified. The matrix $\delta \textbf{B}$
has three eigenvalues and three eigenvectors, says, $\lambda_{1}$,
$\lambda_{2}$, $\lambda_{3}$ and $\textbf{v}_{1}$, $\textbf{v}_{2}$,
$\textbf{v}_{3}$, respectively. When the index is not equal to zero,
the null would be a 3D isolated singular point. This means the null
will not fall to a 2D null (according to the definition of 3D
isolated singular point), hence the matrix $\delta \textbf{B}$ is
regular and none of the eigenvalue is zero. The trace of $\delta
\textbf{B}$ is equal to the sum of the three eigenvalues. On the
other hand, the trace can be presented by
$\bigtriangledown\bullet\textbf{B}$ which is zero in a magnetic
field. The three eigenvalues may be all real numbers or one is real
number, the other two are conjugated complexes. However, the real
parts of them should be (+ - -) or (- + +) (because the sum of the
three real parts is zero), if they are all real numbers, the null is
type $A$ or type $B$, if two of them are complexes, the null is type
$A_{s}$ or type $B_{s}$ (Cowley 1973; Lau \& Finn 1990). One
eigenvector ($\textbf{v}_{1}$ for example), according to the
eigenvalue which is the only positive or negative, points out the
direction of $\gamma$ line (or spine) and the other two
($\textbf{v}_{2}$ and $\textbf{v}_{3}$) determine $\Sigma$ surface
(or fan) near the null-point (Fukao et al. 1975; Lau \& Finn 1990).
In practice, we select a small sphere (radius equals to one grid
unit) including each null (cause the sphere is big enough to include
the cell). The spines and fan should intersect the sphere at two
points and a circle respectively (Fig.~3). The vectors in the two
points would parallel to the spine eigenvector and the vectors on
the circle would parallel to the plane determined by the two fan
eigenvectors. Therefore we can know which point on the sphere
belongs to the intersection of the line of spines or fan by check
the directions of the vectors on the sphere whether parallel to the
directions of spine or fan. We draw lines of fan departing from the
sphere along the circle (denote $l(x)$, where $x$ is belong to the
circle), if a infinite little distance $\epsilon$ in point $x$ make
the other endpoints of $l(x-\epsilon)$ and $l(x+\epsilon)$ sharply
depart from each other, then there must be a separator line
$l_{s}(x)$ start from $x$. When the two endpoints of a separator are
both null points, the separator is called a null-null line. Finally
we obtain the magnetic skeleton which is composed of nulls, spines
fans and separators.

\section{Topology skeleton of AR 10720}
\label{sect:skeleton}

In brief summary using the observed vector magnetograms as
boundary conditions, we first reconstruct the 3D magnetic fields
in the AR's atmosphere with a quasi-linear force-free code (Wang
et al. 2001); then we identify the 3D magnetic nulls by a
differential geometry method (Greene 1992; Zhao et al. 2005) in
the extrapolated vector fields; finally we determine the magnetic
skeleton by connecting the nulls, spines, fans and separators. We
compare the skeletons with flare ribbons and EIT light structures,
also we examine the skeleton evolution from the derived
time-sequence of topologies.

 \begin{figure}
 \centering
   \includegraphics[scale=0.5]{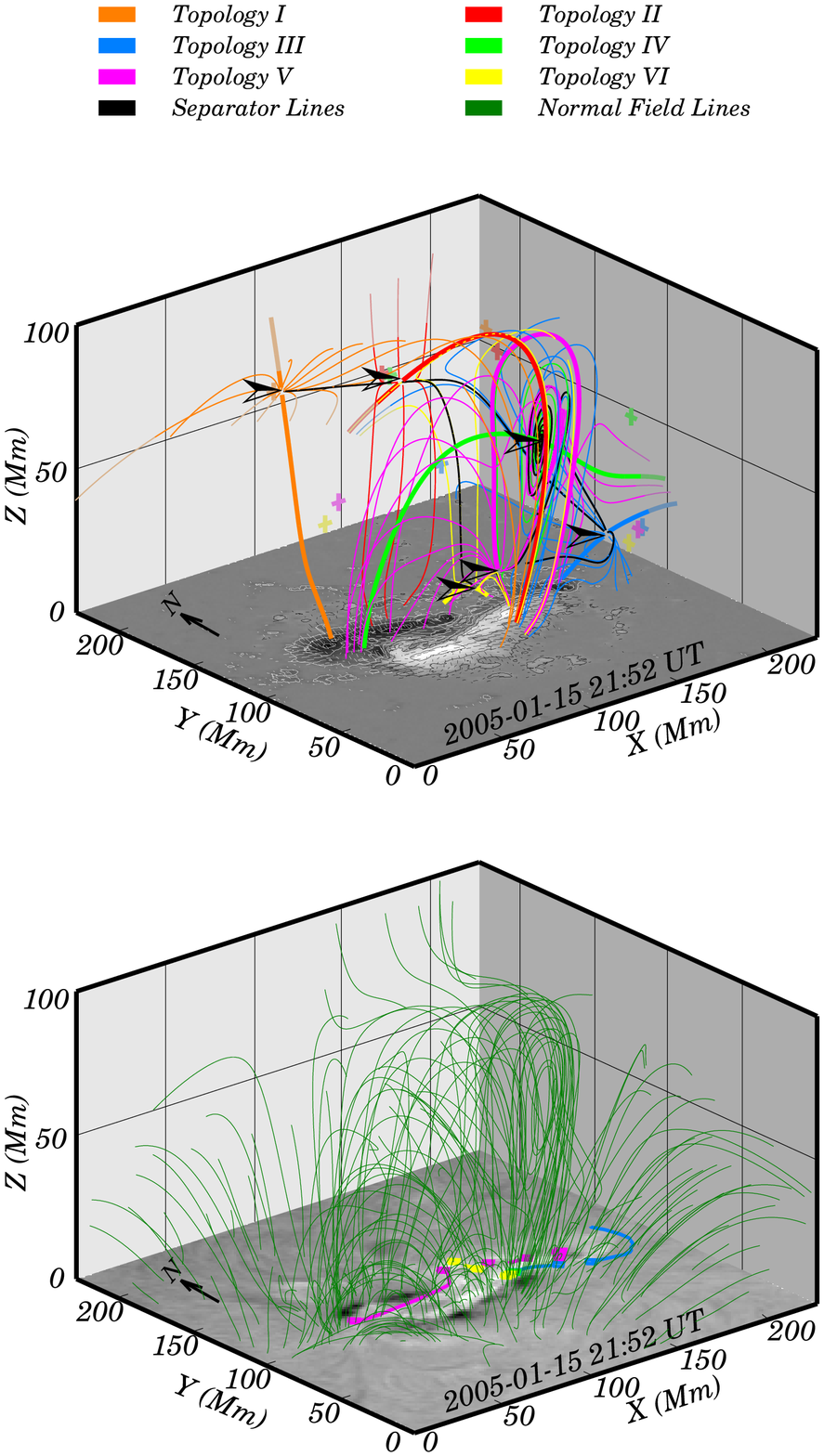}
   \caption{The 3D magnetic structures and determined topology skeleton of
AR10720. The magnetic lines of force extrapolated by a quasi-linear
force-free code based on the observed BBSO vector magnetogram at
21:52 UT of January 15, 2005 is shown in lower panel; while the
magnetic skeleton determined from the extrapolated 3D fields is
shown in upper panel. The vector magnetogram is displayed as the
background of upper one, in which the line-of-sight flux density is
shown by grey map with contours of 200, 700, 1200 Gauss and bright
(dark) color for positive (negative) flux; the transverse
magnetogram is shown by short arrows with length representing the
field strength. Each null and its spines and fan is referred to as a
topology, and numbered from `I' to `VI'. The positions of the nulls
are indicated by arrows while their projections in the two vertical
planes are marked by symbols `+'. Different topologies are drawn
with distinct colors. Thick lines are spines while thin lines are
magnetic lines on the fans. A BBSO H$\alpha$  filtergram is shown as
background of panel lower one, in which the intersections of fans
with chromosphere are drawn to show their correlation with flare
ribbons.    }
   \label{Fig:skeleton}
  \end{figure}

\begin{table}[]
  \caption[]{Information of nulls at 21:52 UT.We employ the coordinate
   in Figure~1 to present the
nulls' positions. $\triangle B$=100\,G (0.01\,T), L=1 unit grid
(1.523\,Mm). $|\langle \triangle B \rangle /\langle L \rangle|$
denotes the average error in the derivatives of the observed
magnetic field.}
  \label{tb1}
  \begin{center}\begin{tabular}{ccccc}
  \hline\noalign{\smallskip}
No  &  Type    & Position     &   Eigenvalues     &     $\frac{|\nabla\bullet B|}{|<\triangle B>/<L>|}$ \\
    &          &   (Mm)     &  (10$^{-10}$ T~m$^{-1}$)  &  \\
  \hline\noalign{\smallskip}
I   & $A$      & (111,222,55) & 0.967, --0.115, --0.194                     & 0.010    \\  
II  & $B$      & (172,215,49) & --0.167, 0.080, 0.592                      & 0.008    \\  
III & $A$      & (207,117,11) & 1.847, --0.523, --0.814                     & 0.008    \\  
IV  & $A_{s}$  & (177,125,47) & 0.308, --0.038$\pm$0.735\textbf{\emph{i}}  & 0.004    \\
V   & $B$      & (148,120,9)  & --3.120, 1.079, 4.405                      & 0.036    \\
VI  & $A$      & (140,126,3)  & 4.795, --2.620, --4.738                     & 0.039    \\
  \noalign{\smallskip}\hline
  \end{tabular}\end{center}
\end{table}

\subsection{Determined Topology Skeleton}
The determined skeleton at 21:25 UT of January 15 basically consists
of six nulls (see Fig.~4). Each null and its associated
$\gamma$-lines (spines) and $\Sigma$-surface (fan) are referred to
as a topology, and numbered as Topologies I, II, ..., VI,
respectively. Their locations, eigenvalues, and divergence of
magnetic vector in term of observational errors are listed in
Table~1. Majority of the magnetic lines of force (see the lower
panel of Fig.~4) which anchored in the photosphere are bounded by a
combination of fans and spines of the Topologies I, II, III, and V,
forming AR's magnetized atmosphere. Topology IV is embedded in the
middle of the atmosphere.

Amongst the six, Null IV is of $A_{s}$ type, i.e., a spiraling-in
null. It locates at an altitude of 47 Mm. For the first time we have
identified a spiral null in the Sun's corona of an AR, based on
observations. The spine of $A_{s}$ null lies above the whole length
of the magnetic neutral line, along which the 'sheared EFRs' emerged
(see Figs.~1 and 2). Around the spine of the $A_{s}$ null the
magnetic lines of force twisted, forming a magnetic wreath shaped
like trumpet shell, i.e., a complicated flux rope structure. Around
the spine of the $A_{s}$ null the magnetic lines of force twisted,
forming a magnetic wreath shaped like trumpet shell, i.e., a
complicated flux rope structure. This structure obviously can be a
kind of physical entities of flux rope, a concept commonly used in
solar physics, but a rope structure may not have an $A_{s}$ or
$B_{s}$ null in its center. The Fan V and a part of the Fan III join
to the twisted lines of force. In fact, the rotated lines of force
which consist the loop-like structure just boundary with Fan III,
Fan IV and Fan V. The electric current density flowing along the
spine is as high as 3$\times$10$^{-4}$\,A~m$^{-2}$ at such an
altitude in the corona. It is, at least, two times higher than that
in surroundings, indicating a concentration of free magnetic energy.
The currents also flew in the fan surface, which resulted in a tilt
angle of 75 degrees between the spine and fan.

The main loop-like structure is consisted by the Topology IV and
Topology V, which derived from Null IV and Null V, respectively. As
we know, the complicated structure centered with a null-null pairs
which include one or two spiral nulls is only described by numerical
simulations based on plasma kinetic approaches (B\"{u}chner 1999;
Cai et al. 2006). For demonstrating the $A_{s}$-$B$ null pairs'
structure more clearly, we simplified the structure in ideal
condition by a cartoon in Figure 5. The $\Sigma_{A_{s}}$ intersects
the $\Sigma_{B}$ with $A_{s}$-$B$ null-null line, the separator, and
the $\Sigma_{A_{s}}$, $\Sigma_{B}$ are semi-infinite sheets bounded
by the $\gamma_{B}$, $\gamma_{A_{s}}$, respectively. The field lines
on $\Sigma_{A_{s}}$ twisted into the $A_{s}$ null, and $\Sigma_{B}$
rolls over onto $\gamma_{A_{s}}$ (note that the part of the field
lines on the $\Sigma_{B}$ asymptotically approach the
$\gamma_{A_{s}}$ but can not touch it). The field lines on the
$\Sigma_{B}$ which asymptotically approach the $\gamma_{A_{s}}$ and
the field lines on the $\Sigma_{A_{s}}$ compose a loop-like
structure. We should point out that in Figure~4 the Fan IV, a
$\Sigma_{A_{s}}$ surface, is not a semi-infinite sheet because it is
bounded not only by Spine V, a $\gamma_{B}$ line, but also by
photosphere.

The six nulls are connected by a net of null-null lines, the
so-called separators, upon which the frozen-in magnetized plasma
of four individual topologies closely contacts and interacts one
another. The adjacent null pairs in the separators are all of
$A$-$B$ or $A_{s}$-$B$ type. These fans and their intersection,
the separators, separate the magnetized atmosphere into different
domains in which the lines of force have the same connectivity.
The net of separators bridging distinct topologies goes through
the magnetized atmosphere. This fact hints the characteristic of
3D magnetic reconnection in the AR. As the reconnection could take
place successively or simultaneously along the whole separator
net, the magnetic energy release during the reconnection in the AR
should be global, fast and explosive in nature (Priest et al.
1997; Priest \& Forbes 2000; Parnell 2007).

\begin{figure}
    \centering
   \includegraphics[scale=0.5]{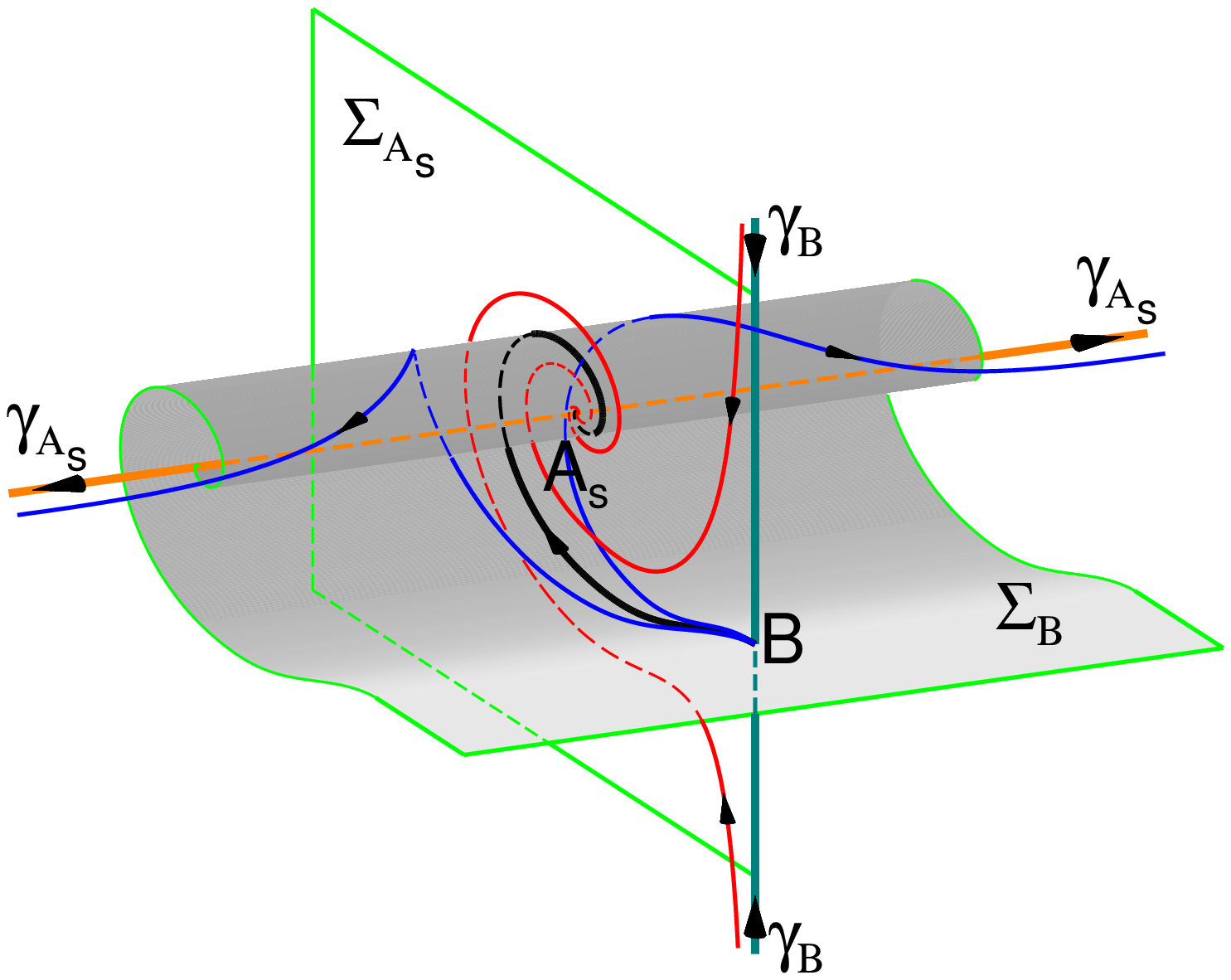}
   \caption{ A cartoon used to simplify the topology structure of the
$A_{s}$-$B$ null-null pair. Orange and celadon lines are
$\gamma_{A_{s}}$ and $\gamma_{B}$ lines respectively. Red and blue
lines present the fields lines on $\Sigma_{A_{s}}$ and
$\Sigma_{B}$. The black line at which $\Sigma_{A_{s}}$ and
$\Sigma_{B}$ intersect is the $A_{s}$-$B$ null-null line. Note,
the twisting sense of magnetic lines of force on both sides of the
$\Sigma_{A_{s}}$ is opposite. }
   \label{Fig:As-B}
\end{figure}

\subsection{Spacial Relations with Flare Ribbons and Brightening EUV
Loops}

It is interesting to see that the cross-sections of spans of the
lower Topologies III, V, and VI are coincided with the main parts of
the flare ribbons (see lower panel of Fig.~4) as described by many
authors previously (see D\'{e}moulin et al. 1997). The south-west
piece of flare ribbons seemed to apparently avoid from all topology
structures, but, in fact, was clearly connected to the fans of Null
V by a set of magnetic arcades straddling the magnetic neutral line.

Such complicated rope structures organized by a spiral null were
never anticipated until we draw out the topology skeleton. The Fans
IV and V intersect at the $A_{s}$-$B$ null-null line, the separator;
they are bounded asymptotically by Spines V and IV, respectively.
The field lines on Fan IV twisted into the Null IV, the $A_{s}$ type
null, and Fan V rolls over onto Spine IV. The field lines on the Fan
V, which asymptotically approach the Spine IV, and field lines on
the Fan VI and a fraction of Fan III compose the frame of the
magnetic wreath. It is remarkable that the magnetic lines of force
on both sides of Fan IV have opposite twisting, i.e., the magnetic
helicity in the rope-like structure does not keep the same sign.
Although the current data analysis does not reveal the details of
associated 3D reconnection, we witness the close similarity between
the 'magnetic wreath' and EUV rope-like structures, brightened in
the course of the flare/CME (see Fig.~6). The EUV brightening took
place in a twisting structure similar to the magnetic lines of force
in the magnetic wreath, said, the general shape and sharp boundaries
of EUV structure well resembled the magnetic wreath. This implies
the close correlation between the energy release and topology
structure.
\begin{figure}
   \centering
   \includegraphics[scale=0.5]{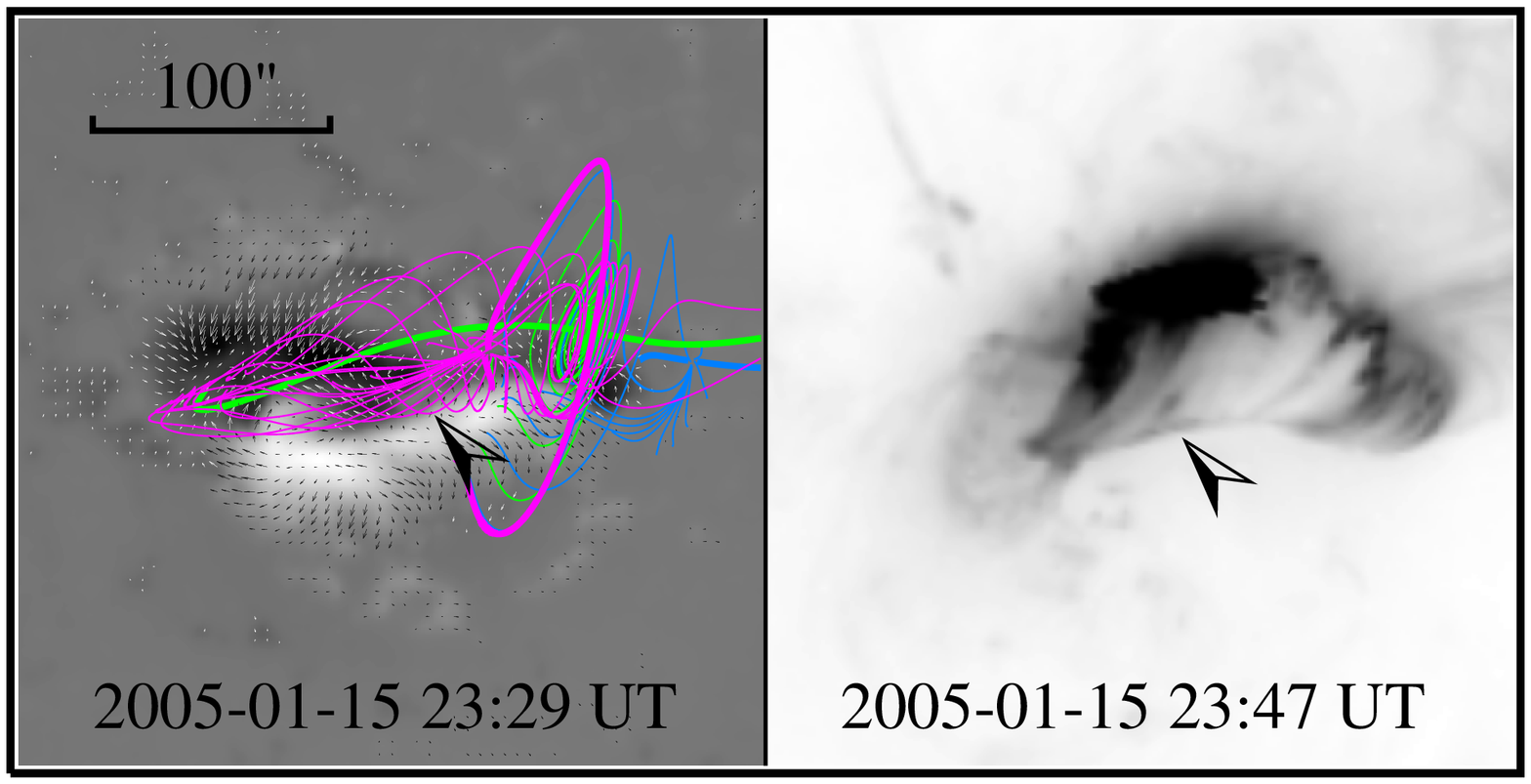}
   \caption{ Comparison of the magnetic wreath associated with the spiral
null (left panel) and EUV brightening structure in the late phase of
the X2.6 flare (right panel). Left: The topology skeleton is
superposed above a BBSO vector magnetogram at 23:29 UT. Right: The
general appearance of the EUV brightening structures is shown by the
reversed color table in which the brightest structures are darkest.
The EUV structures resemble the topology skeleton, particularly the
low sharp boundary seems to coincide with the low boundary of the
wreathed $\Sigma$ surfaces of Nulls III and V (the two boundaries
are pointed by arrows).}
   \label{Fig:EIT loop}
\end{figure}

\begin{figure}
   \centering
    \includegraphics[scale=0.55]{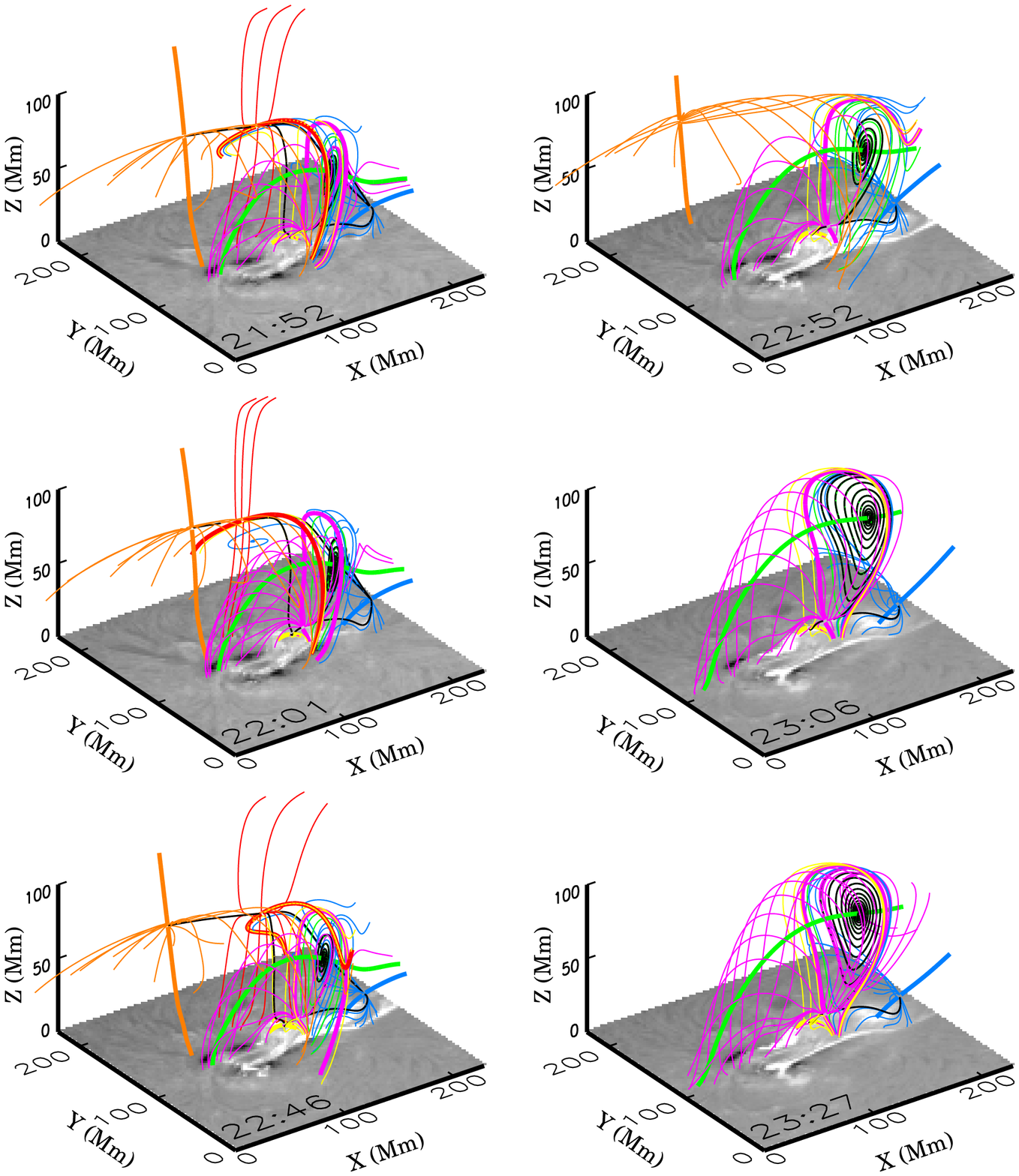}
   \caption{ Skeleton evolution in the interval from pre-state to the
recovery phase of the flare/CME. Null II and its topology
disappeared between 22:46 to 22:52 UT, in which the CME initiated
from the solar disk; while the null I and its topology disappeared
between 23:03 to 23:06 UT at the maximum phase of the flare. The
drawing style of this figure is same as Fig.~1. The elevation of
null IV and the mounting up of magnetic wreath are clearly shown in
the time-sequence of skeleton evolution. }
   \label{Fig:skeleton evolution}
  \end{figure}

\subsection{Skeleton Evolution}

Based on the time sequence of observed vector magnetograms, we
further deduced the skeleton evolution from 21:52 to 23:47 UT (see
Fig.~7), covering the interval from pre-status to recover-phase of
the flare/CME. In the course of the activity event, the topology
skeleton showed obvious evolution, which could be categorized into
slowly systematic evolutions and explosive eruption. One of the key
systematic evolutions is elevation of the spiral null together with
its spines. The spiral null moved up more than 26\,Mm during the
studied interval. At the same time, the magnetic wreath grew. In
accordance, there is also an expansion of Topology I. In the
expansion, Null I moved northward more than 50\,Mm from 21:52 to
22:52 UT. The dramatic evolution is likely a response to the rapid
growth of the `sheared EFRs'. During this interval the apparent
separation speed of opposite polarity flux of the EFRs was about
1.0\,km~s$^{-1}$. The fast and continuous separation of opposite
polarities of the new emerging flux can be seen clearly from the
south-west movement of the negative flux patch (already grew to a
sunspot) indicated by a thick arrow in Figure~2.

In the course of the systematic evolution two topology eruptions
took place in association with the flare-CME development. Null II
and its topology erupted right before the flare and shortly after
the initiation of the halo CME at 22:36 UT, which passed into LASCO
C2 field of view at 23:06 UT with a speed of more than
2800\,km~s$^{-1}$. The second topology eruption appeared at the
maximum phase of the flare and manifested as the disappearance of
Null I and its topology. The separators connecting the Nulls I and
II disappeared in the two topology eruptions, which is indicative of
the vigorous 3D magnetic reconnection along these null-null lines in
the flare/CME event. However, we may not exclude a possibility that
the two `disappeared' topologies might simply move outside the
calculation domain. To test this possibility, we have extended our
calculations to more than 200\,Mm above the photosphere, but found
no hints of their existence in the higher corona. During the
skeleton evolution, the three lower nulls, Nulls III, V, and VI, and
their topologies remained, keeping the spatial relations with flare
ribbons unchanged.

\section{Discussion}
\label{sect:discussion}

There may appear some uncertainties in determination of magnetic
skeleton, which come mainly from the 3D reconstruction of magnetic
fields. We have examined the reliability of the approaches in this
work by artificially introducing random noises in the observed
magnetograms, and found that the basic skeletons that we
determined are re-producible.

It is well-known that with the same boundary condition different
algebra and codes would give different force-free solutions, i.e.,
the extrapolation of force-free fields from observations is not
unique (Seehafer, 1978). We have tested the robusticity of our
findings by using different force-free codes. First of all, the
potential and linear force-free field extrapolation models and
methods (Alissandrakis 1981; Seehafer, 1978; Chiu and Hilton,
1977) could not well reproduce the skeleton which we determined.
The potential and exactly constant-$\alpha$ force-free
extrapolations were imbecilic in finding spiral nulls. The reason
is that opposite helicity signs maintain in the magnetic
structures when there is a spiral null, and the co-existence of
opposite sign helicity, hence the $\alpha$ value, is not allowed
by potential or linear force-free models.

These new skeletons reconstructed by other extrapolation codes
have rare relation with the flare ribbons and it revealed that the
potential and exactly constant-$\alpha$ force-free extrapolations
were imbecilic in finding spiral nulls and a right transverse
field is necessary for reconstruction based on a complicated
vector magnetogram with strong shearing.

For severely twisted or sheared magnetic topology, the codes based
on boundary element method (BEM) which are developed by Wang et
al. (2001) and Yan \& Sakurai (2000) may have some advantage. In
the comparison based on an analytical model (Low \& Lou 1990), the
BEM (Liu et al. 2002) performed better than potential and constant
$\alpha$ method and has less disparity nearer the center of the
extrapolation box (Schrijver et al. 2006).

We have been trying to use other non-linear force-free models
(Wheatland et al. 2000; Wheatland 2006, 2007; Song et al. 2006) to
examine if the current results can be repeated by other code
independently. More careful work still needs to be done before
drawing definitive conclusions. While the complicated topology
skeleton determined in this study has not been reproduced by other
potential and linear force-free codes so far, the reality of the
determined skeleton can only be hinted by its comparison with
observations of brightness structures in H$\alpha$, EUV, and X-ray
wavebands. As shown by Figures~4, 6 and 7, the H$\alpha$ flare
ribbons are closely associated with the lower nulls and their
topologies, while the higher nulls (including the spiral null) and
their topologies are more closely correlated to the CME onset.
Because our determined topology skeleton does match some of the
activity structures in both spatial and temporal domains, we believe
that our determination represents, at least, some aspects of the
real 3D topology peculiarity.

More active regions are under consideration and a few different
force-free codes are undertaken as future efforts to
observationally determine the topology skeleton of magnetic fields
in active regions.

In the case that the 3D vector field measurements, either by
\emph{in situ} or remote-sensing, are available, our method to
determine the 3D nulls is robust and ready to be used.

\section{Conclusion}
\label{sect:conclusion}

Based on the time sequence of observed vector magnetograms with
high resolution and sensitivity, we have determined the 3D
magnetic skeleton and its evolution in a solar AR. We have
identified, for the first time, the spiral magnetic null in the
Sun's corona. Extremely complicated rope structures, the magnetic
wreath, associated with the spiral null appear to be the central
ingredient of the magnetic skeleton. The determined skeleton has
close tempo-spatial associations with solar explosive activity.
The skeleton evolves in responds to the vector field evolution in
the solar photosphere. We have clearly uncovered two topology
eruptions in the course of the flare/CME. The complexity of
observed magnetic skeleton and its evolution in solar active
atmosphere was not previously reported in the literature.

Although spiral nulls are predicted in theoretical approaches (Lau
\& Finn 1990; Parnell et al. 1996), they have never been discovered
in solar observations. However, their appearance and role in 3D
magnetic reconnection have been clearly described by numerical
simulations based on plasma kinetic approaches (B\"{u}chner 1999;
Cai et al. 2006). Our observations fully support the theoretical
predictions and numerical simulations. For many years solar
astronomers have hypothesized the common existence of flux ropes in
Sun's corona. The flux rope concept becomes a central element in CME
modelling in solar physics. However, the observational
identifications were not available in the literature. Different
authors use the same scientific term, `flux rope', but refer to
different physical entities in observations (Hudson \& Cliver 2001).
Here we propose that the magnetic topology of a spiral null in the
corona, e.g., the magnetic wreath, serves, at least, one type of the
real flux ropes. Flux ropes are likely to have spiral nulls inside
in reality. Their magnetic structures appear to be more complicated
than that one has thought. Within a flux rope opposite signs of
magnetic helicity can be kept in association with topology
peculiarity. The magnetic reconnection between opposite helicity
flux may result in explosive release of more free magnetic energy
(Wang et al. 2004)

\begin{acknowledgements}
This work was supported by National Basic Research Program of
China (2006CB806303) and National Natural Science Foundations of
China (G10573025 and 40674081). We should thank for the help of H.
N. Wang, E. R. Priest, M. S. Wheatland and M. T. Song.
\end{acknowledgements}

\label{lastpage}


\begin{thebibliography}{99}
\small \setlength{\itemindent}{-3mm}
\setlength{\itemsep}{-0.5mm}
\setlength{\baselineskip}{4.5mm}

  \bibitem[1986]{ai86} Ai G., Hu Y., 1986, Publ. Beijing Astron. Obs., 8, 1

  \bibitem[2001]{aul81} Alissandrakis C. E., 1981, A\&A, 100, 197

  \bibitem[2000]{aul14} Aulanier G., DeLuca E. E., Antiochos S. K. et al., 2000, ApJ, 540, 1126

  \bibitem[2001]{bro7} Brown D. S., Priest E. R., 2001, A\&A, 367, 339

  \bibitem[1999]{buc24} B\"{u}chner J., 1999, Astrophys. and Space Sci., 264, 25

  \bibitem[1996]{bun96} Bungey T. N., Titov V. S., Priest E. R., 1996, A\&A, 308, 223

  \bibitem[2006]{cai25} Cai D., Nishikawa K., Lembege B., 2006, Plasma Phys. Control. Fusion, 48, B123-B135

  \bibitem[1977]{chi77} Chiu Y. T., Hilton H. H., 1977, ApJ, 212, 873

  \bibitem[1973]{cow27} Cowley S. W. H., 1973, Radio Sci., 8, 903

  \bibitem[1992]{dem10} D\'{e}moulin P., Henoux J. C., Mandrini C. H., 1992, Solar Phys., 139, 105

  \bibitem[1996]{dem96} D\'{e}moulin P., Henoux J. C., Priest E. R., Mandrini C. H., 1996, A\&A, 308, 643

  \bibitem[1997]{dem97} D\'{e}moulin P., Bagal$\acute{a}$ L. G., Mandrini C. H. et al.,
                1997, A\&A, 325, 305

  \bibitem[1999]{fil13} Filippov B., 1999, Solar Phys., 185, 297

  \bibitem[2001]{fle15} Fletcher L. et al., 2001, ApJ, 554, 451

  \bibitem[1975]{fuk75} Fukao S., Masayuki U., Takao T., 1975, Rep. Ionos. Space Res. Jpn., 29, 133

  \bibitem[1946]{gio1} Giowanelli R. G., 1946, Nature, 158, 81

  \bibitem[1988]{gor88} Gorbachev  V. S., Somov B. V., 1988, Solar Phys., 117, 77

  \bibitem[1992]{gre21} Greene J. M., 1992, J. Comp. Phys., 98, 194

  \bibitem[2001]{hud26} Hudson H. S., Cliver E. W., 2001, J. Geophys. Res., 106, 25199

  \bibitem[1993]{lau6} Lau Y. T., 1993, Solar Phys., 148, 301

  \bibitem[1990]{lau3} Lau Y. T., Finn J. M., 1990, ApJ, 350, 672

  \bibitem[2006]{li17} Li H., Schmieder B., Aulanier G., Berlicki A., 2006, Solar Phys., 237, 85

  \bibitem[2002]{liu02} Liu Y., Zhao X. P., Hoeksema J. T. et al., 2002, Solar Phys., 206, 333

  \bibitem[2005]{lon12} Longcope D. W., 2005, Living Rev. Solar Phys., 2, 7

  \bibitem[2002]{lon11} Longcope D. W., Klapper I., 2002, ApJ, 579, 468

  \bibitem[1987]{low87} Low B. C., 1987, ApJ, 323, 358

  \bibitem[1990]{low90} Low B. C., Lou Y. Q., 1990, ApJ, 352, 343

  \bibitem[1988]{low88} Low B. C., Wolfson R., 1988, ApJ, 324, 574

  \bibitem[2003]{mai16} Maia D. et al., 2003, A\&A, 405, 313

  \bibitem[2007]{par8} Parnell C. E., 2007, Mem. S.A.It., 78, 229

  \bibitem[1996]{par4} Parnell C. E., Smith J. M., Neukirch T., Priest E. R., 1996,J. Plasmas Phys., 3, 759

  \bibitem[1997]{pri5} Priest E. R., Bungey T. N., Titov V. S., 1997, Geophys. Astrophys. Fluid Dyn., 84, 127

  \bibitem[1995]{pri95} Priest E. R., D\'{e}moulin P., 1995, J. Geophys. Res., 100, 23,443

  \bibitem[2000]{pri23} Priest E., Forbes T., 2000, Magnetic reconnection: MHD theory and applications,
                   Cambridge: Cambridge Univ. Press

  \bibitem[2006]{sch06} Schrijver C. J., Derosa M. L., Metcalf T. R. et al., 2006, Solar Phys., 235, 161

  \bibitem[2001]{see78} Seehafer N., 1978, Solar Phys., 58, 215

  \bibitem[1986]{see9} Seehafer N., 1986, Solar Phys., 105, 223

  \bibitem[2005]{spi19} Spirocks T., 2005, thesis, New Jersey Institute of Technology

  \bibitem[2006]{son06} Song M. T., Fang C., Tang Y. H. et al., 2006, ApJ, 649,1084

  \bibitem[1958]{swe2} Sweet P. A., 1958, IAU symposium, 6, 123

  \bibitem[2002]{tit02} Titov V. S., Hornig G., D\'{e}moulin P., 2002, J. Geophys. Res., 107, 1164

  \bibitem[1993]{tit93} Titov V. S., Priest E. R., D\'{e}moulin P., 1993, A\&A, 276, 564

  \bibitem[2001]{wan20} Wang H. N., Yan Y. H., Sakurai T., 2001, Solar Phys., 201, 323

  \bibitem[1996]{wan96} Wang H. N., Wnag J. X., 1996, A\&A, 313, 285

  \bibitem[1996]{wjx96} Wang J. X., Shi Z. X., Wang H. N., L\"{u} Y. P., 1996, ApJ, 456, 861

  \bibitem[2004]{wjx04} Wang J. X., Zhou G. P., Zhang J., 2004, ApJ, 615, 1021

  \bibitem[2000]{whe00} Wheatland M. S., Sturrock P. A., Roumeliotis G., 2000, ApJ, 540, 1150

  \bibitem[2006]{whe06} Wheatland M. S., 2006, Solar Phys., 238, 29

  \bibitem[2007]{whe07} Wheatland M. S., 2007, Solar Phys., submitted

  \bibitem[2006]{xia18} Xiao C. J. et al., 2006, Nature Phys., 2, 478, DOI:
  10.1038/nphys342

  \bibitem[2000]{yan00} Yan Y. H., Sakurai T., 2000, Solar Phys., 195, 89

  \bibitem[2005]{zha22} Zhao H., Wang J. X., Zhang J., Xiao C. J., 2005, \chjaa, 5, 443

\end{thebibliography}
\end{document}